\documentclass[amsmath,amssymb,letterpaper,aps,prl,twocolumn,showpacs]{revtex4}
\usepackage{graphicx}
\usepackage[usenames]{color}
\usepackage[latin1]{inputenc}

\begin{document}

\title{Magnetic exchange mechanism for electronic gap opening in graphene}
 \author{T. G. Rappoport$^{1}$, M. Godoy$^{1,2}$, B. Uchoa$^{3}$, R. R dos Santos$^{1}$ and A.~H. Castro Neto$^{4}$}

\affiliation{$^{1}$Instituto de F\'{\i}sica, Universidade Federal do Rio de Janeiro, CP 68.528, 21941-972 Rio de Janeiro RJ, Brazil} 

\affiliation{$^{2}$Instituto de F\'{\i}sica, Universidade Federal de Mato Grosso, 78060-900,
Cuiab\'a MT, Brazil.}

\affiliation{$^{3}$Department of Physics$\mbox{,}$ University of Illinois at Urbana-Champaign$\mbox{,}\,$1110
W. Green St, Urbana, IL, 61801, USA}

\affiliation{$^{4}$Department of Physics, Boston University, 590 Commonwealth Avenue, Boston, MA 02215, USA}

\date{Version 11.1 -- \today}
\begin{abstract}
We show within a local self-consistent mean-field treatment that a random distribution of magnetic adatoms can open a robust gap in the electronic spectrum of graphene. The electronic gap results from the localization of the charge carriers that arises from the interplay between the graphene sublattice structure and the exchange interaction between the adatoms. The size of the gap depends on the strength of the exchange interaction between carriers and localized spins and can be controlled by both temperature and external magnetic field. 
 Furthermore, we show that an external magnetic field creates an imbalance of spin-up and spin-down carriers at the Fermi level, making doped graphene suitable for spin injection and other spintronic applications.

\end{abstract}
\maketitle

Graphene is a two-dimensional sheet of carbon atoms whose elementary
excitations are massless Dirac fermions \cite{graphene}. Due to the vanishing density of states (DOS) near the Dirac points, 
 adatoms, such as H,  which are otherwise non-magnetic and bond strongly to carbon atoms, can easily form local magnetic moments in graphene \cite{uchoa08}. The combination of properties such as ballistic transport, extremely large
 spin coherence lengths \cite{injection1}, and gating effects makes this material very attractive for spintronic
applications \cite{zutic04}. 

For certain applications, such as in the fabrication of transistors,
it would be desirable to open a large gap in graphene. 
However, opening a gap in suspended samples has 
turned out to be unexpectedly difficult. 
 Dirac fermions have long inelastic mean free paths and are typically insensitive to disorder and localization effects. The Dirac points seem also very robust against
instabilities in general.  Localization effects, nevertheless, can be also triggered by scattering mechanisms that operate in the spin channel, which is not very sensitive to the relativistic nature of the electrons. Due to the presence of sublattice quantum numbers, when magnetic adatoms are adsorbed on top of graphene,  the interaction between two local spins siting on opposite sublattices is antiferromagnetic, whereas spins on the same sublattice interact ferromagnetically \cite{saremi07,brey07,annica10,sherafati11}. The interplay of magnetic interactions and disorder can give rise to the formation of spin textures and clusters~\cite{rappoport09}, which can localize charge carriers as a result of the exchange interaction between local and itinerant spins. 
In this Letter, we show that the adsorption of magnetic
adatoms constitutes a surprisingly flexible way to localize charge carriers and open a controllable gap in graphene. 

We address this issue by considering a tight-binding model, in which local magnetic moments are randomly distributed on a honeycomb lattice, and interact with the spin of the itinerant electrons through a local Heisenberg exchange interaction; this effective interaction results from the interplay between hybridization effects and Coulomb interactions, similarly to the Kondo Hamiltonian for metals \cite{Kondo}. We assume a self-consistent semiclassical picture where the local spins are treated classically but the itinerant electrons are treated quantum mechanically, accounting for the realization of the disorder in the localized spins~\cite{berciu01,berciu04}.  In the numerical results, we show the emergence of a large localization gap that is driven by the  concentration of adatoms and the exchange coupling, and which can be controlled by temperature and external magnetic field. Furthermore, when the chemical potential is shifted above or below the gap, we propose that the application of an external magnetic field can spin polarize the charge carriers at the Fermi level, making doped graphene useful for spintronics applications.
 Although this effective description does not account for the change of the energy spectrum due to 
the covalent character of the bonding and lattice distortions caused by the incorporation of the adatoms~\cite{chem}, this model is justified when the magnetic exchange is the dominant interaction. In graphene, the exchange coupling of the Kondo Hamiltonian has been predicted to be as large as a few eV \cite{uchoa10,kotliar10}.

We start from a random distribution of magnetic impurities (or spins) sitting on top of carbon atoms. For spins on top carbon sites, the exchange Hamiltonian in graphene has a simple form \cite{uchoa10}, and the total Hamiltonian, including charge carriers and external fields, is 
\begin{eqnarray}
{\cal H}&=&-t\sum_{\langle i,j\rangle}^{}(a^{\dagger}_{i\sigma} b_{j\sigma}+h.c)
+\sum_{i}^{} J_{i} \vec{S}_i\cdot\vec{s}_i \nonumber \\
&&- g \mu_BH\sum_{i}^{}s^z_i - \tilde{g} \mu_B H \sum_{i}^{}S^z_i,
\label{1}
\end{eqnarray}
where $a_{i\sigma}$ ($b_{i\sigma}$) annihilates an electron with spin $\sigma$  on site $i$ of sublattice $\alpha=A$ ($B$).
The first term in Eq. (\ref{1}) describes charge carriers hopping between nearest neighbor graphene sites, with $t\simeq$ 2.7 eV  %\cite{graphene}, 
while the second one couples localized ($\vec{S}_i$) and itinerant ($\vec{s}_i$) spins through the local exchange constant $J_i$ $> 0$ ; the last two terms in Eq. (\ref{1}) describe the interaction of the carriers and spins with an external {\em in-plane} magnetic field $H$. 
Although the exchange interaction is local, the indirect interaction among the localized spins is long ranged, which, in principle, allows the formation of a true ordered state in two dimensions, in spite of  thermal fluctuations. 

At the mean-field level,
$
\vec{S}_i\cdot \vec{s}_{i} \rightarrow S(i)
s^z_i + S^z_i s(i) - S(i)s(i),
$
where the magnetizations $S(i)\equiv \langle S^z_i \rangle$ and $s(i) \equiv \langle {
s}^z_i \rangle $ 
must be computed self-consistently at each individual site ~\cite{berciu01,berciu04}. 
The mean-field decomposition results in an effective Hamiltonian of charge carriers  that can be fully diagonalized  for any configuration of localized spins. We then calculate the expectation value of the carrier spins $s(j)$ on all lattice sites. The expectation value of each {\em local} spin can be computed back by enforcing $S(i)$ to be equal to the local magnetization due to charge carriers at each impurity site, in the presence of the external field, $S(i)={\cal B}_S(\beta H_i)$, where $H_i=J \langle {s}^z_i \rangle-\tilde{g} \mu_B H$ and ${\cal B}_S(x)$ is the Brillouin function. The process is iterated until self-consistency is achieved for all localized spins \cite{berciu01}. As a result, for a given random configuration of localized spins (with concentration $n_s$) and fixed density of carriers $n_e$, we determine the chemical potential, $\mu$, the energy, $E_{n\sigma}$, on the $n$-th level and the corresponding probability amplitude, $|\psi_{n\alpha,\sigma}(j)|^2$, for occupation of a carrier state with spin $\sigma$ on a given site $j$ on sublattice $\alpha$. With these values, we extract the magnetizations of the localized and itinerant spins on each sublattice, $M_\alpha$ and $m_\alpha$, respectively, and the DOS.  The classical 
spins $S(i)$ range from 0 to an arbitrary value $S_{max}$. Here we set $S_{max} = 1/2$.

We have performed calculations for systems with up to $N$=$2\times 65\times 65$  sites, and used random distributions of magnetic impurities corresponding to a given spin concentration $n_s$; from now on, all quantities should be understood as configurational averages taken in ensembles varying between 10 and 50 disorder realizations, which suffice to yield small enough error bars.  As we are interested in the case where the Fermi level is close to the Dirac point, in the following we set $n_e=1$ (one electron per carbon). 

\begin{figure}[t]
\includegraphics[width=0.65\columnwidth,clip]{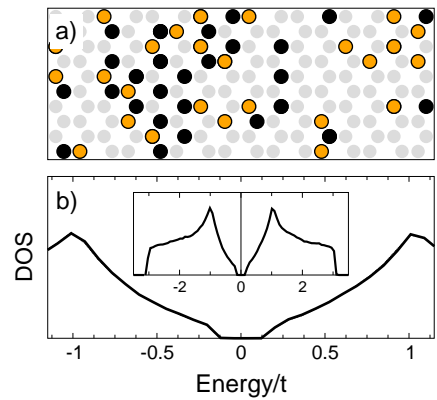}
\caption{(Color online) 
(a) Typical random spin configuration (spin
up in black, down in orange) for $n_s =0.4$, $N = 2 \times 65 \times 65$:  ($\mu = 0$); (b) density of states (DOS) averaged over 
disorder realizations represented by panel (a). The inset in panel (b) shows the DOS over a wider energy range. Vertical lines indicate
the Fermi level. All data are for $ J = t$, and $k_B T = 0.002t$.} 

\label{fig1}
\end{figure}

As a check of our scheme, we verified that calculations for full coverage of local moments ($n_s = 1$) at  $\mu=0$  correctly reproduce the AFM ground state predicted by perturbative methods \cite{saremi07,brey07}; 
at low temperatures, our results also agree with those obtained through Monte-Carlo simulations with frozen charge degrees of freedom~\cite{rappoport09}.
For  a random distribution of impurities, Fig.~\ref{fig1}(a) shows one localized spin configuration for $n_s<1$ and $\mu=0$; we see that both FM and AFM clusters are formed. The corresponding carrier DOS is depicted in  Fig~\ref{fig1}(b): a gap clearly opens around the Dirac point \cite{moreo10,note}. 

%%%%%%%%%%%%%%%%%%%%%%%%%%%%%%%%%%%%%%%%%%%%%%%%
%%%%%%%%%%%%%%%%%%%%%%%%%%%%%%%%%%%%%%%%%%%%%%%%

\begin{figure}[t]
\includegraphics[width=0.9\columnwidth,clip]{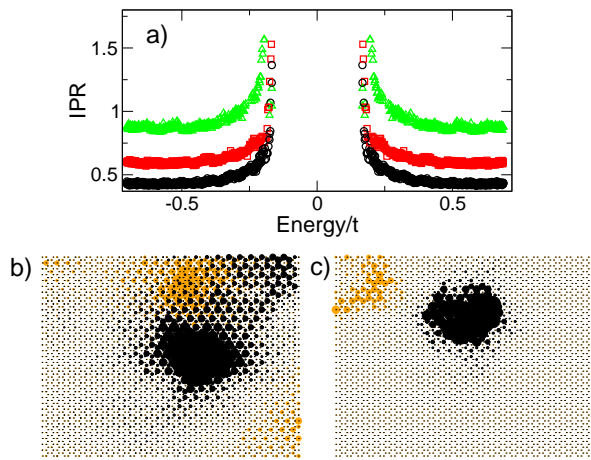}
\caption{(Color online) 
(a) Inverse participation ratio (IPR) calculated under the same conditions as in Fig 1(a)
for three system sizes:  $N = 4050$ (green triangles),
$N = 6050$ (red squares) and $N = 8450$ (black circles);
(b)-(c) Typical eigenstate with energy in the vicinity of the gap for  $J=t$ (b) and $J=2t$ (c). 
The size of the symbol is proportional to the relative weight of the eigenstate in a given site (spin up in black, down in orange). Dark and orange regions represent spin polarized droplets (see text).}
\label{fig2}
\end{figure}

The mechanism of gap opening in this material strongly resembles the double-exchange mechanism in manganites~\cite{manganites}. 
Hopping does not flip the electron spin while it moves from one sublattice to another;  since neighboring adatoms (occupying different sublattices) are antiferromagnetically aligned, it costs an energy $J$ for an electron to move between sublattices, leading to electronic localization. For isolated adatoms, the average energy cost, and hence the localization length, will scale with the dilution. In Fig. \ref{fig2}(a), we show the inverse participation ratio (IPR), which measures the degree of localization of states, and is defined by the quantity $P_{\sigma n}=\sum_{i\alpha}^{N} |\psi_{n\alpha,\sigma}(i)|^4$. 
For an extended state, $P_{\sigma n}$ scales with $1/N$,  
whereas for a true localized state it does not scale with the system size, and its IPR is comparatively larger.  
States near the gap have a significantly larger IPR than those with  energies away from it.
Therefore, the former states are localized, while the latter, whose IPR clearly scales with $N$, are extended. 
In the dilute limit, these localized states resemble self-trapped magnetic polarons, similar to the ones proposed by Nagaev~\cite{Nagaev01}:  carriers are localized by exchange interaction in spin-polarized droplets located at  energetically favorable positions, whose sizes scale with $1/J$ (see Fig~\ref{fig2} (b)-(c) ). Our numerics also indicates that the localization of the charge carriers is robust against local gating effects induced by the local reconstruction of the electronic density at the impurity sites. We considered this effect by adding a diagonal term in the graphene Hamiltonian that changes the on-site energy at sites where the magnetic ions are attached. We find that the gap is still present for on-site energies $\lesssim J$. Since the exchange coupling  can be extremely large \cite{uchoa10,kotliar10}, the localization effects discussed here may be dominant for a variety of adatoms in graphene.

%%%%%%%%%%%%%%%%%%%%%%%%%%%%%%%%%%%%%%%%%%%%%%%%
%%%%%%%%%%%%%%%%%%%%%%%%%%%%%%%%%%%%%%%%%%%%%%%%

\begin{figure}[b]
\includegraphics[width=1.0\columnwidth,clip]{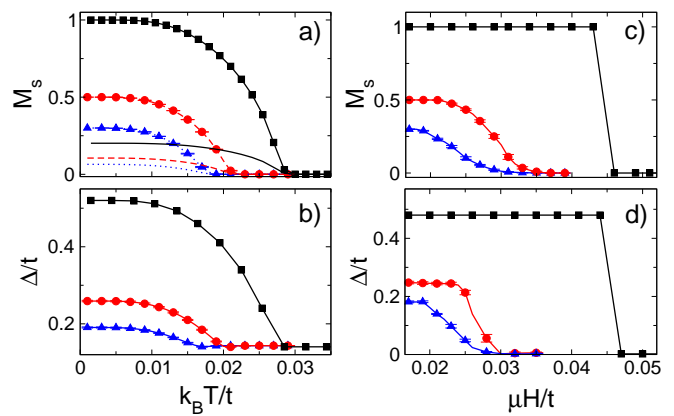}
\caption{(Color online) 
(a) Staggered magnetization versus temperature for localized spins and carriers and different impurity concentrations: $n_s=1$ (black squares for $M_s$, and full line with no symbols for $m_s$), $n_s=0.5$ (red circles and dashed line) and $n_s=0.3$ (blue triangles and dotted line);
(b) Gap ($\Delta$) as a function of the temperature [the color code for $n_s$ in (a) is the same in all panels]; 
(c) Staggered magnetization as a function of applied magnetic field $H$;
(d) Gap as a function of $H$.  $J=0.5t$, $\mu=0$ and $N=2\times50 \times 50$ in all cases.}
\label{fig3}
\end{figure}

We first examine the behavior of the staggered magnetizations, $M_s=|M_A-M_B|/S$ and $m_s=|m_A-m_B|/s$, corresponding to local and itinerant spins, respectively. From Fig.~\ref{fig3}(a), we see that due to the nature of the RKKY interaction in graphene, even for a diluted system with ferromagnetic clusters [see, e.g., Fig~\ref{fig1}(a)], $M_s$ attains its maximum value, $n_s$, at low temperatures. 
The spins of carriers and magnetic impurities are antiferromagnetically coupled to each other by the exchange interaction, and hence both magnetizations vanish at the same critical temperature $T_c$, for a given concentration of adatoms ($n_s$).
Figure \ref{fig3}(b) shows the temperature dependence of the gap, $\Delta(T)$, from which the correlation between the curves in panels (a) and (b) is clear: $\Delta$ decreases from its saturated value (proportional to  $n_sJ$~\cite{moreo10} )  at low $T$, tracking the behavior of both $M_s$ and $m_s$, until it reaches a minimum value when $M_s=m_s\simeq 0$; 
due to the finiteness of the system, this minimum value is not zero, hence setting a minimum energy scale.
Below $T_c$, the resistivity of the system displays an activated behavior, typical of a semiconductor for $k_BT\ll\Delta(T=0)$, but undergoes a metal-insulator (MI) transition at intermediate temperatures. 
The strong temperature dependence of the gap and the temperature-induced MI transition resemble the transport characteristics of Kondo insulators~\cite{risebor00}.

%%%%%%%%%%%%%%%%%%%%%%%%%%%%%%%%%%%%%%%%%%%%%%%%
%%%%%%%%%%%%%%%%%%%%%%%%%%%%%%%%%%%%%%%%%%%%%%%%

When a sufficiently strong magnetic field $H_c$ is applied, aligning the local spins, electronic hopping is no longer hindered: the charge carriers delocalize, and the system becomes metallic. 
This behavior can be seen from Figs~\ref{fig3}(c) and (d), which illustrate that for a fixed $T$, the gap also tracks the behavior of $M_s$ with $H$: $\Delta$ vanishes when the spins in both sublattices point in the same direction. 
This dependence has important consequences for the transport properties of magnetically doped graphene: the system will display a very strong negative magnetoresistance, since $H$ tends to close the gap and to increase dramatically the conductivity of the system. 
This effect is very similar to the one observed in manganites and in magnetic semiconductors, such as EuSe~\cite{manganites} and magnetic hexaborides \cite{vitor04}; it could also have bearings on the colossal negative magnetoresistance observed in dilute fluorinated graphene \cite{hong10}.

%%%%%%%%%%%%%%%%%%%%%%%%%%%%%%%%%%%%%%%%%%%%%%%%
\begin{figure}[b]
\includegraphics[width=1.0\columnwidth,clip]{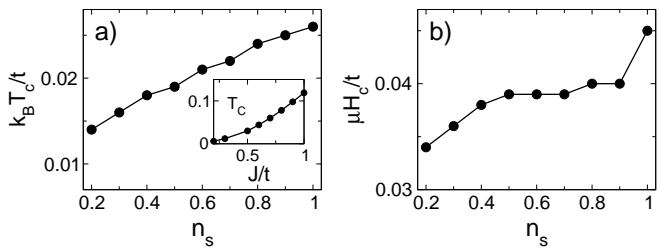}
\caption{(a) Critical temperature $T_c$ ($M_s=0$) versus spin concentration $n_s$. (b) Critical field $T_c$ versus $n_s$ for $T=0.001t$. $J=0.5t$ $N=2\times 36\times 36$
% and all values are in units of $t$. 
The inset in panel (a) shows $T_c$ in function of $J$.}
\label{fig4}
\end{figure}
%%%%%%%%%%%%%%%%%%%%%%%%%%%%%%%%%%%%%%%%%%%%%%%%

As it can be seen from Fig.~\ref{fig3}, magnetic field, temperature, and concentration of magnetic adatoms can be used to control the magnitude of the localization gap in graphene. % since it tracks the staggered magnetization $M_s$.  
In order to manipulate $\Delta$, it is necessary to establish the range of parameters for which the gap opens or closes. 
To this end, we have varied the adatom concentration, $n_s$, and determined the critical temperature, $T_c$, and the critical field, $H_c$, for fixed temperature.
%; some of the results are summarized in Fig.~\ref{fig3}. 
The critical lines in the main panel of Fig.\ref{fig3}(a) scale quadratically with $J$,  $T_c\propto J^2$, for $n_s=1$, as seen from the inset of Fig.~\ref{fig4}(a).  
 As one increases the adatom concentration from $n_s=0.2$, the carrier polarization tends to increase, which, in turn, enhances the correlation among the local moments. As expected, $T_c$ and $H_c$ rise monotonically with the amount of disorder [see Fig. \ref{fig4}(b)].

\begin{figure}[t]
\includegraphics[width=1\columnwidth]{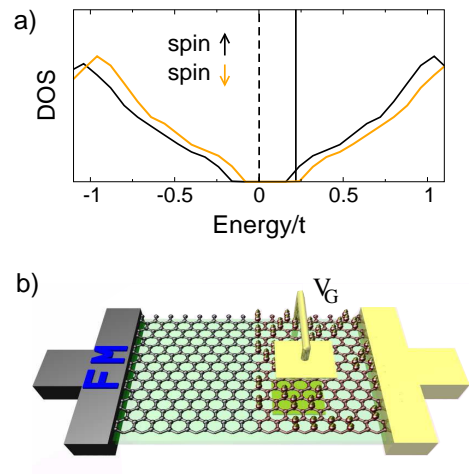}
\caption{(a) Electronic density of states at finite magnetic field for $J=t$, $n_s=0.5$, $\mu H=0.12t$, $k_BT=0.001t$. Dashed vertical line: $\mu=0$, corresponding to an insulating state; solid vertical line: $\mu=0.2t$ (metallic). Black (dark) curve represents spin up and orange (light) curve, spin down. (b) A spin injection device (see text).}
\label{fig5}
\end{figure}

 In the numerics, we find that both the energy gap and the DOS are not strongly dependent of $\mu$. Since the external magnetic field breaks the degeneracy between spin-up and spin-down states, a controlled variation of $\mu$, caused e.g., by a gate voltage,  can drive the system from an insulator to a metallic regime with spin-polarized carriers, where only spin-up (or spin-down) states are available. This is illustrated in Fig.~\ref{fig5}(a), where we depict the DOS at finite magnetic field  for $\mu=0$ (dashed vertical line), when the system is an insulator, and for $\mu \simeq 0.2 t$ (solid vertical line), when it becomes metallic with a spin polarized density of charge carriers at the Fermi level. Hence, the application of a external gate voltage,  $V_G$,  can be used to change the spin polarization of the states at the Fermi level, permitting the generation and control of spin-polarized currents through the use of randomly distributed adatoms in graphene.

Noting that a single graphene layer can be separated in different regions, with and without the presence of adatoms, if during the deposition process part of the layer is covered with a mask, one can produce and test spin polarized currents using a device sketched in Fig.~\ref{fig5}(b). Half of the layer is covered with magnetic adatoms, with which a non-magnetic (normal) contact is made, while the clean part has a magnetic contact, such as Cobalt. An external field is turned on and produces the spin polarized shift in DOS, while the gate voltage moves $\mu$ with respect to the Dirac point [Fig.~\ref{fig5}(a)]. 
If a current passes through the system from the normal to the magnetic contact, spin-polarized currents are generated in the magnetically doped part of graphene and injected in the clean part, where it will be tested by the magnetic contact. Differently from other spin-injection schemes experimentally tested~\cite{injection1}, in this case injection has no external agents, thus eliminating unwanted sources of scattering such as interface mismatch between graphene and other materials, like magnetic metals.

%%%%%%%%%%%%%%%%%%%%%%%%%%%%%%%%%%%%%%%%%%%%%%%%
%%%%%%%%%%%%%%%%%%%%%%%%%%%%%%%%%%%%%%%%%%%%%%%%

These ideas can also be used to propose a spintronics device similar to a spin-valve~\cite{baibich88}. Instead of separating the layer in two parts, as above, it is now separated in three: two magnetically doped parts separated by a clean one. The doped parts are subject to independent $V_G$'s and both have normal metallic contacts.  
With the same $H$ and different $V_G$'s, each magnetic part can have a specific spin imbalance at the Fermi Level. If a current passes through the system in both directions, the magnetoresistance can be controlled by the $V_G$'s .
The device also produces spin-polarized carriers in one side, and tests the polarization in the other.  
%In the calculations leading to Fig. 4(a), we have used large values of $J$ and $n$  for illustrative purposes only. 
 In the device sketched in Fig. 5 (b),  for a spin concentration of $n_s = 0.1$, and exchange coupling as low as $J = 0.1$ eV, spin polarized carriers can be produced by fields less than 1T, combined with gate voltages of the order of just a few mV. 

In conclusion,  we have established that a random distribution of magnetic adatoms can localize charge carriers and open a gap in graphene. 
The mechanism of gap opening is the hindering of hopping due to the antiferromagnetic correlation of  magnetic impurities on opposite sublattices. 
The size of the gap depends on the concentration of adatoms, and on the strength of the exchange coupling; it can therefore be controlled by temperature and external magnetic field. 
These two adjustable parameters can drive graphene into a metal-insulator phase transition in the presence of magnetic disorder. Furthermore, we showed that a magnetic field can be used to produce spin polarized carriers, suitable for spin injection and other spintronic applications.

We thank  E. Fradkin for discussions. 
The research by TGR, MG, and RRdS was supported by the Brazilian Agencies CNPq, CAPES, and FAPERJ; MG also acknowledges partial support from FAPEMAT. 
AHCN acknowledges the ONR grant MURI N00014-09-1-1063, and BU acknowledges partial support from the DOI grant DE-FG02-91ER45439 at University of Illinois.


\begin{thebibliography}{99}
\bibitem{note}  If the spins are forced to occupy only one sublattice, the ground state is FM but the sublattice symmetry is broken, leading to a gap in the electronic excitation spectrum for each spin.  
The spin degeneracy is also lifted and the corresponding DOS for spin up and spin down carriers are shifted in opposite directions, leading to a metallic state.
\end{thebibliography}
\end{document}